\documentclass[preprint2]{aastex}

\shorttitle{SuperWASP observations of XO-1b}
\shortauthors{Wilson et al.}

\begin{document}

%% LaTeX will automatically break titles if they run longer than
%% one line. However, you may use \\ to force a line break if
%% you desire.

\title{SuperWASP Observations of the Transiting Extrasolar planet XO-1b}

%% Use \author, \affil, and the \and command to format
%% author and affiliation information.
%% Note that \email has replaced the old \authoremail command
%% from AASTeX v4.0. You can use \email to mark an email address
%% anywhere in the paper, not just in the front matter.
%% As in the title, use \\ to force line breaks.

\author{D. M. Wilson\altaffilmark{1}, B. Enoch\altaffilmark{2}, D. J. Christian\altaffilmark{3}, W. I. Clarkson\altaffilmark{2,4}, A. Collier Cameron\altaffilmark{5}, H. J. Deeg\altaffilmark{6}, A. Evans\altaffilmark{1}, C. A. Haswell\altaffilmark{2}, C. Hellier\altaffilmark{1}, S. T. Hodgkin\altaffilmark{7}, K. Horne\altaffilmark{5}, J. Irwin\altaffilmark{7}, S. R. Kane\altaffilmark{5,8}, T. A. Lister\altaffilmark{5,1}, P. F. L. Maxted\altaffilmark{1}, A. J. Norton\altaffilmark{2}, D. Pollacco\altaffilmark{3}, I. Skillen\altaffilmark{9}, R. A. Street\altaffilmark{3}, R. G. West\altaffilmark{10}, P. J. Wheatley\altaffilmark{11}}

\altaffiltext{1}{Astrophysics Group, School of Chemistry and Physics, Keele University, Staffordshire, ST5 5BG, UK}
\altaffiltext{2}{Department of Physics and Astronomy, The Open University, Milton Keynes, MK7 6AA, UK}
\altaffiltext{3}{Astrophysics Research Centre, Main Physics Building, School of Mathematics and Physics, Queen's University, University Road, Belfast, BT7 1NN, UK}
\altaffiltext{4}{Space Telescope Science Institute (STScI), 3700 San Martin Drive, Baltimore, MD 21218, USA}
\altaffiltext{5}{School of Physics and Astronomy, University of St. Andrews, North Haugh, St. Andrews, Fife, KY16 9SS, UK}
\altaffiltext{6}{Instituto de Astrof\'{i}sica de Canarias, C/V\'{i}a L\'{a}ctea, s/n, E-38200 La Laguna, Tenerife, Spain }
\altaffiltext{7}{Institute of Astronomy, University of Cambridge, Madingley Road, Cambridge, CB3 0HA, UK}
\altaffiltext{8}{Department of Astronomy, University of Florida, 211 Bryant Space Science Center, Gainesville, FL 32611-2055, USA}
\altaffiltext{9}{Isaac Newton Group of Telescopes, Apartado de correos 321, E-38700 Santa Cruz de la Palma, Tenerife, Spain}
\altaffiltext{10}{Department of Physics and Astronomy, University of Leicester, Leicester, LE1 7RH, UK}
\altaffiltext{11}{Department of Physics, University of Warwick, Coventry CV4 7AL, UK}

%% Mark off your abstract in the ``abstract'' environment. In the manuscript
%% style, abstract will output a Received/Accepted line after the
%% title and affiliation information. No date will appear since the author
%% does not have this information. The dates will be filled in by the
%% editorial office after submission.

\begin{abstract}
We report on observations of 11 transit events of the transiting extrasolar planet XO-1b by the SuperWASP-North observatory. From our data, obtained during May-September 2004, we find that the XO-1b orbital period is 3.941634$\pm$0.000137 days, the planetary radius is 1.34$\pm$0.12 $R\rm_{Jup}$ and the inclination is 88.92$\pm$1.04$^{\circ}$, in good agreement with previously published values. We tabulate the transit timings from 2004 SuperWASP and XO data, which are the earliest obtained for XO-1b, and which will therefore be useful for future investigations of timing variations caused by additional perturbing planets. We also present an ephemeris for the transits.
\end{abstract}


\section{Introduction}

Although the current list of confirmed exoplanets stands at over 180\footnote{http://exoplanet.eu/} only 10 have been found to transit their parent star. Transiting planets allow the true planetary mass, radius, density and inclination to be determined allowing us to place constraints on fundamental theories of planet formation and evolution. 
The SuperWASP project is one of a number of wide-angle searches for extra-solar planetary transits. In this paper we demonstrate the capabilities of the SuperWASP cameras and transit-searching algorithm by reporting observations of 11 transits of an 11th magnitude star by the exoplanet XO-1b \citep{xo}, observed during the first season of SuperWASP observations in 2004.

\section{SuperWASP Observations}

The SuperWASP Wide Angle Search for Planets project\footnote{http://www.superwasp.org} is an automated ultra-wide angle photometric survey covering both Northern and Southern hemispheres. SuperWASP-North (SW-N) is based on La Palma, Canary Islands, and SuperWASP-South (SW-S) is based at SAAO, South Africa. Both observatories consist of 8 cameras, each with an 11.1cm aperture Canon 200mm f/1.8 lens backed by a 2k$\times$2k EEV CCD. Each camera has a field of view of $7.8\times7.8$ degrees with a 13.7$^{\prime\prime}$/pix plate scale, resulting in a total field of view of almost 500 square degrees per observatory. Further details of the project are given in \citet{pollacco}.

The SW-N observatory obtained nearly 4500 individual observations of XO-1 over 150 days between 2 May 2004 to 29 September 2004. The object was recorded by two cameras producing a total of 8875 measurements. The photometric precision, when outliers from cloudy nights are excluded, is approximately 9mmags (RMS). This is slightly worse than usual for stars of this magnitude owing to the close proximity to the edges of the camera fields.

Lightcurves from SuperWASP are de-trended using the algorithm of \citet{tamuz} before being passed through \textsc{hunter} \citep{acc}, a transit search algorithm based on the method of \citet{protopapas}. The algorithm computes $\chi^2$ values of transit model lightcurves using a box-shaped model slid over the observed lightcurve. Typically, a few tens of transit-like lightcurves are identified from each CCD field of 10-20,000 objects.

\begin{figure}
\begin{center}
\includegraphics[scale=0.5,angle=270]{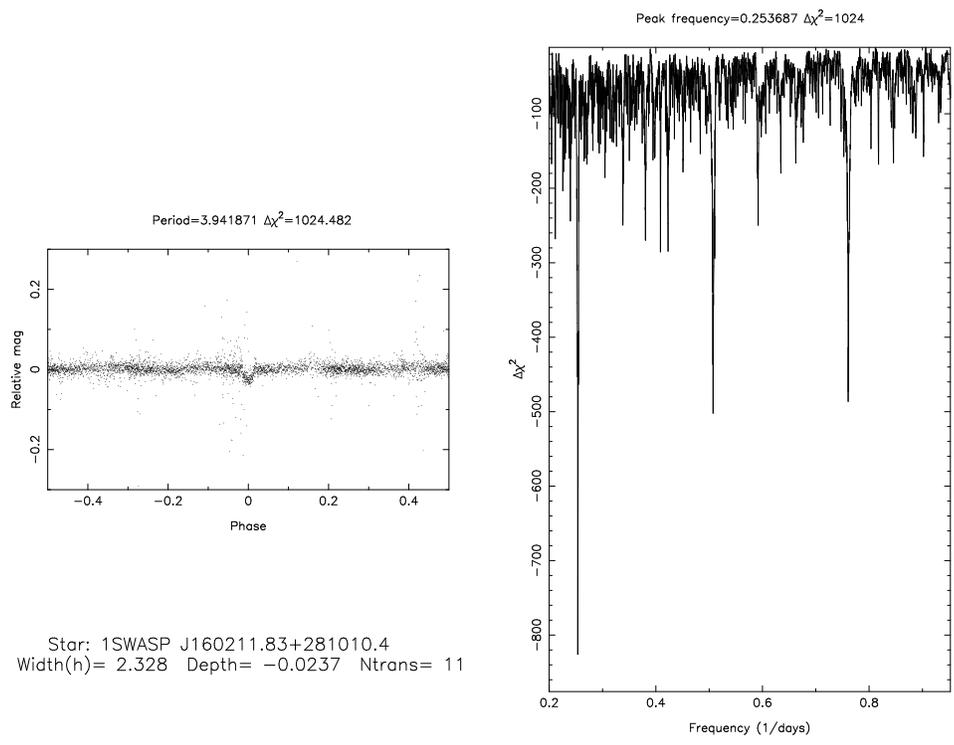}
\figcaption[f1.ps]{Hunter output for object 1SWASPJ160211.83+281010.4 (camera 2) including periodogram and lightcurve folded on the Hunter determined best-fitting period of 3.942134 days.\label{hunter}}
\end{center}
\end{figure}

The \textsc{hunter} output for XO-1 (1SWASP\,J160211.83+281010.4) is shown in Figure \ref{hunter}. The periodogram shows the value of $\chi^2$ for the best least-squares fit at each frequency; the de-trended lightcurve is phase-folded on the best fitting period of 3.942134 days. \textsc{hunter} listed this object as a high priority candidate to be considered for further investigation.

\section{Lightcurve Fitting}

The SW-N data covers 11 transit events and, excluding outliers, consists of 8468 datapoints. The transits were fitted to a simulated planetary transit generated using \textsc{ebop} (Eclipsing Binary Orbit Program; \citet{ebop}). \textsc{ebop} uses biaxial ellipsoids to simulate eclipsing binary star systems; however, by considering the secondary as an opaque disc, a transiting planetary system can easily be modelled. The simulated lightcurve is dependent upon the radii ratio of the transiting planet to the parent star $R\rm_{pl}$/$R_{\star}$, the inclination of the transiting planet's orbit, $i_p$, and the limb-darkening coefficient of the star.

The limb-darkening coefficient was determined by convolving the SuperWASP bandpass with fluxes and monochromatic coefficients listed by \citet{hamme}. A linear limb-darkening coefficient of 0.565 was calculated for the stellar temperature of 5750K (G1V) quoted by \citet{xo}.

The best-fit parameters determined from a least-squares fit to all transits simultaneously are listed in Table~\ref{table:params}. Figure 2 shows the data phase-folded on the best-fitting period with the best-fit model overplotted. The original XO survey data (Peter McCullough, private communication), obtained on a similar instrument to SW-N, are also shown for comparison. The errors were generated using a boot-strap Monte Carlo method in which we generated and re-fit 1000 simulated data sets from the best-fit lightcurve with the same sampling and noise characteristics as the observed lightcurve. The planetary radius was determined from the ratio $R\rm_{pl}$/$R_{\star}$ by using the stellar radius of 1.0$\pm$0.08$R_{\sun}$ determined spectroscopically by \citet{xo}. The parameters determined from the fit are consistent with previously published values \citep{xo}.

11 transit events were identified from the determined period and ephemeris and the best-fit model was fitted to each of these individually to determine the time of mid-transit $T_0$ (Table~\ref{table:timings}). The best fit model was also fitted to the XO survey data which cover 3 transit events and were also obtained in 2004. Three of the total fourteen transits were rejected either due to insufficient coverage of the transit event, or in one case the data being too noisy to produce an adequate fit. The errors were generated by perturbing $T_0$ so as to increase $\chi^2$ by one and are typically of the order 5-10 mins, which is comparable to the data sampling rate.

This paper has demonstrated that SuperWASP can detect and characterise exoplanet transits and obtain sufficient data to determine an ephemeris. Other candidate exoplanet transits from our 2004 data will be reported in subsequent papers.

\section{Acknowledgements}
The WASP consortium consists of representatives from the Universities of Cambridge (Wide Field Astronomy Unit), Keele, Leicester, The Open University, Queen's University Belfast and St Andrews, along with the Isaac Newton Group (La Palma) and the Instituto de Astrophysic de Canarias (Tenerife). The SuperWASP-N and SuperWASP-S Cameras were constructed and operated with funds made available from Consortium Universities and PPARC.

We are grateful to the XO team for making available the original XO photometry.

\clearpage

\begin{table}
\caption{Best fitting parameters of XO-1 from the fit of the SW-N data.\label{table:params}}
\begin{tabular}{lr}
\tableline\tableline
Parameter & SW-N\\
\tableline
Period (days) & 3.941634$\pm$0.000137\\
T$_0$ (HJD) & 2453150.6849$\pm$0.0018\\
$R\rm_{pl}$/$R_{\star}$ & 0.138$\pm$0.020\\
$i_P$ (degrees)&  88.92$\pm$1.04\\
$R_{\star}$ ($R_{\sun}$) & 1.0$\pm$0.08\tablenotemark{a}\\
$R\rm_{pl}$ ($R\rm_{Jup}$) & 1.34$\pm$0.12\\
\tableline\tableline
\end{tabular}
\tablenotetext{a}{From \citet{xo}}
\end{table}
\clearpage

\begin{table}
\caption{Best-fit times of mid-transit for full and partial XO-1b transits from SW-N and XO survey data.\label{table:timings}}
\begin{tabular}{llllll}\tableline\tableline
Observatory & HJD (mid-transit) & Transit & N\_Points\tablenotemark{a}\\
\tableline
XO & 2453123 & (no-fit) \tablenotemark{b} &\\
XO & 2453127.0385$\pm$0.0058 & (partial) & 32\\
XO & 2453142.7818$\pm$0.0218 & (partial) & 40\\
SW-N & 2453146 & (no-fit)\tablenotemark{b} &\\
SW-N & 2453150.6855$\pm$0.0106 & (partial) & 88\\
SW-N & 2453154.6250$\pm$0.0026 & (full) & 99\\
SW-N & 2453158.5663$\pm$0.0034 & (full) & 102\\
SW-N & 2453162.5137$\pm$0.0025 & (full) & 117\\
SW-N & 2453166.4505$\pm$0.0025 & (partial) & 68\\
SW-N & 2453170.3917$\pm$0.0037 & (partial) & 65\\
SW-N & 2453229.5143$\pm$0.0045 & (partial) & 54\\
SW-N & 2453233 & (no-fit)\tablenotemark{b} &\\
SW-N & 2453237.4043$\pm$0.0032 & (partial) & 47\\
SW-N & 2453241.3410$\pm$0.0067 & (partial) & 38\\
\tableline\tableline
\end{tabular}
\tablenotetext{a}{Number of data points covering the transit, i.e., within the phases 0.96-1.04}
\tablenotetext{b}{The data for HJD2453123 and HJD2453233 cover less than half of the transit and so were rejected. The data for HJD2453146 are too noisy to provide a reliable transit time measurement.}
\end{table}
\clearpage

\clearpage

\begin{figure}
\plotone{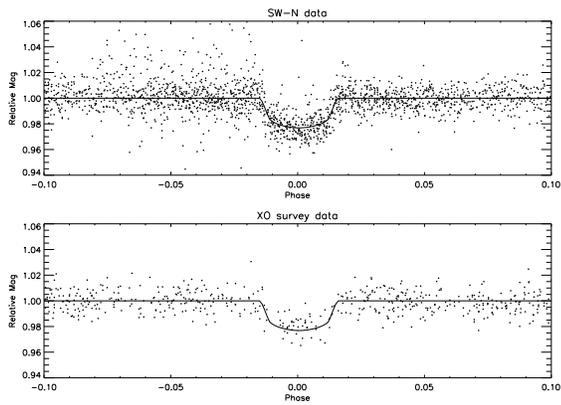}
\figcaption{SuperWASP-North data (upper panel) and XO survey data (lower panel) for XO-1 phase folded on the best fitting parameters listed in Table~\ref{table:params} with the best-fit model over-plotted.\label{swfolded}}
\end{figure}
\clearpage

\end{document}